# A HIGH-Q MICROWAVE MEMS RESONATOR

*Zhu Jian[1], Yu Yuanwei[1], Zhang Yong[1], Chen Chen[1,2], Jia ShiXing[1]*

[1] Nanjing Electronic Devices Institute, Nanjing, 210016, PRC
[2] Dept. of Physics, Nanjing University, 210093, PRC
TEL: 86-25-86858306, email: j-zhu@public1.ptt.js.cn

## ABSTRACT

A High-Q microwave (K band) MEMS resonator is presented, which employs substrate integrated waveguide (SIW) and micromachined via-hole arrays by ICP process. Nonradiation dielectric waveguide (NRD) is formed by metal filled via-hole arrays and grounded planes. The three dimensional (3D) high resistivity silicon substrate filled cavity resonator is fed by current probes using CPW line. This monolithic resonator results in low cost, high performance and easy integration with planar circuits. The measured quality factor is beyond 180 and the resonant frequency is 21GHz.It shows a good agreement with the simulation results. The chip size is only 4.7mm×4.6mm×0.5mm. Finally, as an example of applications, a filter using two SIW resonators is designed.

## 1. INTRODUCTION

The recent development of microwave communication systems has led to drastic constraints on every element of the front-end radio, and particularly on passive elements such as resonators and filters. They have to be smaller size, low cost, to present high quality factor (Q) and to keep planar compatibility in order to easily interconnect them to other elements of the system. However, on the one hand, traditional air cavity waveguide resonator takes up a large circuit area and it can not be compatible with planar circuits although it has a high quality factor (Q). On the other hand, the compatible quarter wavelength (or half wavelength) microstrip (or coplanar waveguide) resonator usually has a low quality factor. A solution is to use Substrate Integrated Waveguide (SIW) techniques [1-2]. The SIW resonator and filter takes the advantages of a proper tradeoff between the size and the performance [3]. Using MEMS (microelectronic mechanical systems) technology, nonradiation dielectric waveguide is formed in the silicon substrate and it results the great size reduction compared with traditional air cavity waveguide. And its quality factor is much higher than the microstrip resonator.

Fig.1 shows the structure of the SIW resonator. The rectangular resonant cavity is built using many rows of via holes in a high resistivity silicon substrate. These vertical via holes are fabricated by DRIE (Deep Reaction Ion Etch) process [4]. The coupling of the cavity with the planar circuit is achieved by an aperture through a common metallic plane. In this paper, we present different coupling structures which lead to different Q of the resonator. Simple design equations for the SIW cavity are described. Simulation and measured results are reported and discussed. Finally, to verify the usefulness of the SIW cavity, a two cavities filter was built. The simple design procedure and simulation results are reported.

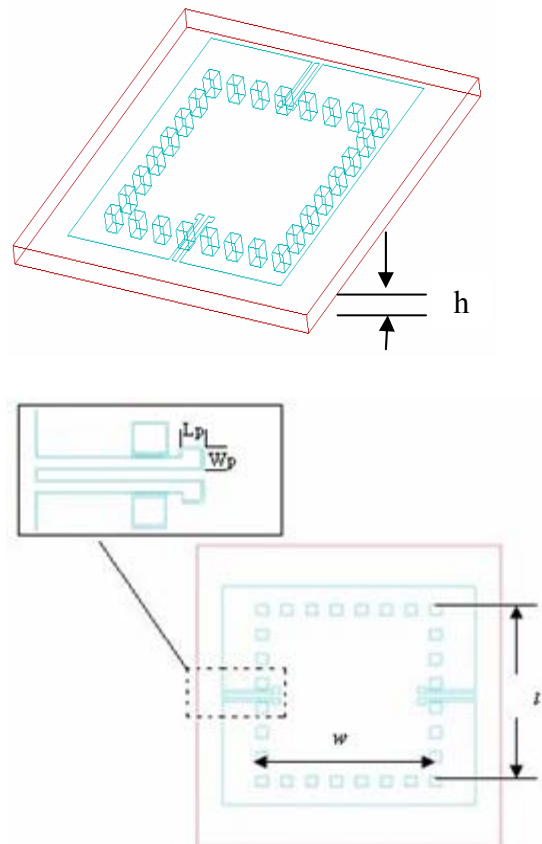

Fig.1.Configuration of the MEMS SIW resonator

## 2. RESONATOR DESIGN

The resonator works in TE101 resonant mode. And the resonant frequency is determined by [5]:

$$f_{101} = \frac{c}{2\pi\sqrt{\mu_r \varepsilon_r}}\sqrt{\left(\frac{\pi}{w_{eff}}\right)^2 + \left(\frac{\pi}{l_{eff}}\right)^2} \quad (E1)$$

Where $w_{eff}$ and $l_{eff}$ are the effective width and length of the SIW cavity, respectively given by

$$w_{eff} = w - \frac{d^2}{0.95p}, \quad l_{eff} = l - \frac{d^2}{0.95p} \quad (E2)$$

The two equations are valid for $p < \lambda_0 \cdot \sqrt{\varepsilon_r}/2$ and $p < 4d$, Where $w$ and $l$ are the width and length of resonant SIW cavity, respectively. d and p are the diameter of metallic via and the space between adjacent vias, respectively. c is the velocity of light in the vacuum. $\mu_r$ and $\varepsilon_r$ are the relative permeability and the relative permittivity of the substrate.

The coupling between the cavity and the CPW planar circuit is made with current probes, as shown in Fig.1. The structure of the coupling probe (determined by Wp and Lp) can influence the Q of the resonator. The unloaded Q of the resonator can be approximated by the following relation [6]:

$$\frac{1}{Q_u} = \frac{1}{Q_L} - \frac{1}{Q_e} \quad (E3)$$

$$Q_L = \frac{f_0}{\Delta f} \quad (E4)$$

$$Q_e = 10^{\frac{-[S_{21}(dB)]}{20}} \quad (E5)$$

where $Q_L$ and $Q_e$ are the loaded Q and the external Q of the resonator. $f_0$ is the resonant frequency, and $\Delta f$ is the 3dB band width.

The resonant frequency designed is 20.5GHz. According to (E1) and the design principles described, d and p are chosen 200um and 250um. The width and length of resonant SIW cavity are calculated: $w$ =3150um, $l$ =3150um.

Once the material of the substrate and the structure of the cavity are decided, the Q is influenced just by Wp, Lp and h.

An electromagnetic (EM) simulation is done. Fig.2 shows the simulation results.

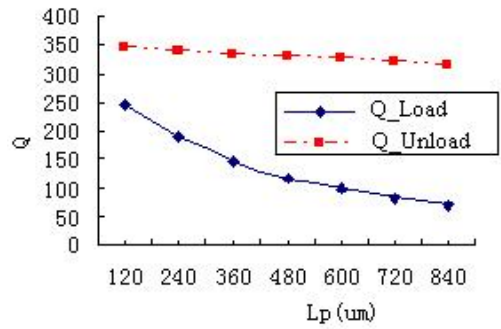
(a) Change Lp

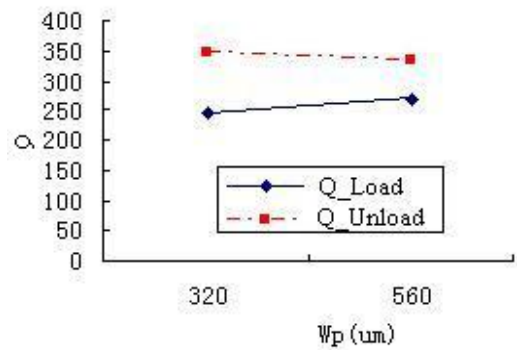
(b) Change Wp

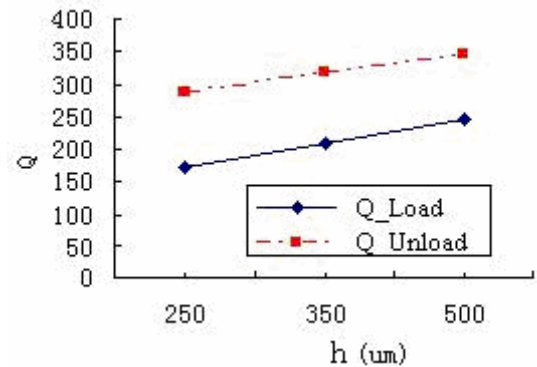
(c) Change h
Fig.2 The simulation Q of the SIW MEMS resonator

According to the analysis, the final values of these parameters are: h=500um, Lp=840um, Wp=320um.

Fig.3 shows the electric field of the resonator, obviously the nonradiation dielectric waveguide is obtained.

## 3. EXPERIMENTAL RESULTS

The resonator is fabricated on the 4″ high resistivity silicon substrate. As shown in Fig.2, large thickness of the substrate will result in high quality factor. However it is difficult to fabricate a deep vertical via-hole. The critical problem is solved by MEMS process. The multistep ICP etching technique is used to get a deep hole with an exactly vertical sidewall that can ensure the good performance of the resonator.

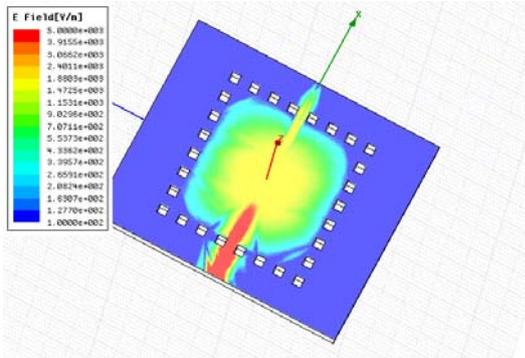

Fig.3 Electric field of the resonator

The final size of the chip is 4.7mm×4.6mm×0.5mm. The sample is shown in Fig.4. And the measured results are presented in Table.1 and Fig.5. The approximation of the resonant frequency is within 3%. This difference is due to the enlargement of the via hole size which is caused by the limited fabrication precision. The measured Qu is around 200, smaller than the simulation results. This is mainly due that the conductor loss and the dielectric loss is higher than the value given by the EM simulation tools.

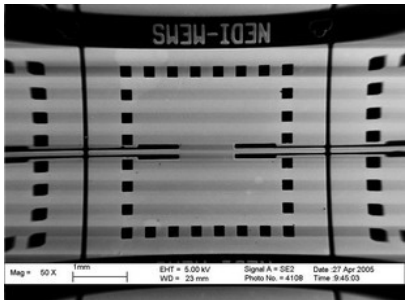

Fig.4. Photograph of the resonator

Table.1 Measured results and simulation results of the resonator.

| | $f_0$(GHz) | Insertion Loss(dB) | Return Loss(dB) | $Q_u$ |
|---|---|---|---|---|
| **Simulation results** | **20.637** | **-2.26** | **-13.8** | **258** |
| Sample 1 | 21.165 | -3.33 | -10.7 | 184 |
| Sample 2 | 21.225 | -2.82 | -10.7 | 202 |
| Sample 3 | 21.150 | -3.20 | -10.7 | 192 |

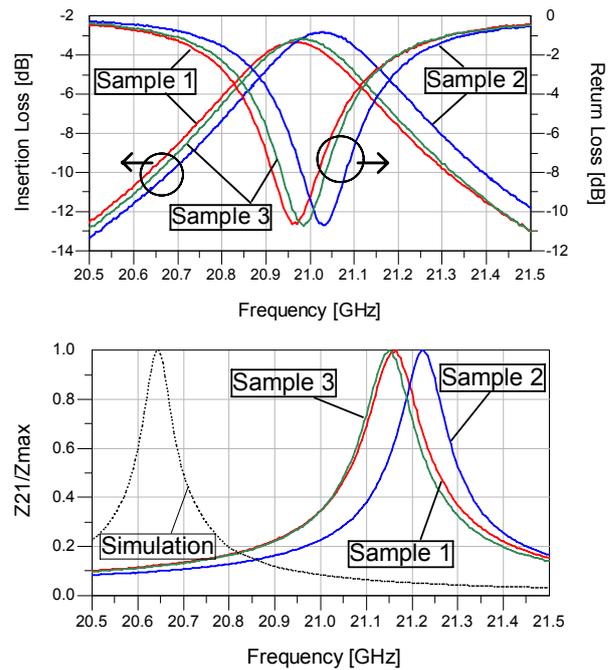

Fig.5 Measured results of 3 samples. (up) S parameter, (down) Z parameter.

### 4. APPLICATION

The SIW resonator can be used in filter, diplexer and oscillator design [7-8]. As an application, a two cavities filter was built. Fig.6 and Fig.7 present the layout and the simulation results. The filter has an insertion loss of 2.0dB at 20.3GHz and a frequency bandwidth of 400MHz (relatively bandwidth of 2%). The die size is only 7mm×4mm. It is significantly that the insertion loss is reduced compared with the traditional microstrip filters and its size is much smaller than the traditional waveguide filters.

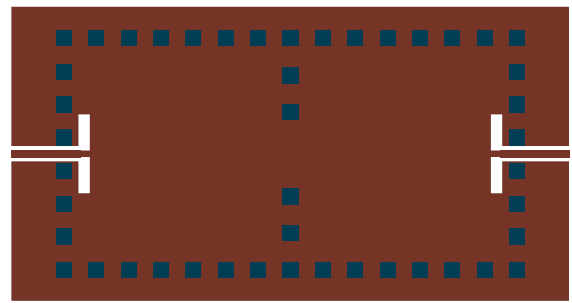

Fig.6. Layout of the filter

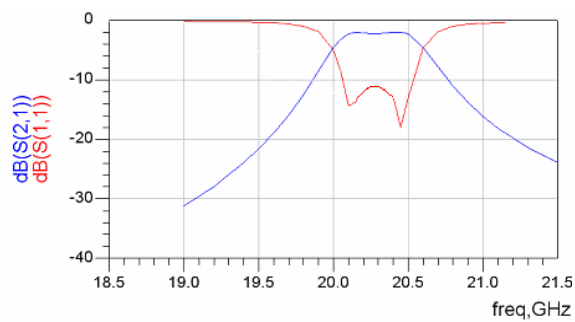

Fig.7. Simulation results of the filter

### 5. CONCLUSION

A MEMS SIW resonator is analyzed and measured. Simple design equations are given. An example of the application in microwave filter is provided. The results show that the SIW resonator has the advantages of high Q and small size. It is certain that this component can be wide used in the microwave filters and oscillators.